\documentclass[aps,prl,preprint,groupedaddress]{revtex4-1}

\usepackage{graphicx}
\usepackage{dcolumn}
\usepackage{bm}
\usepackage{subfigure}

\begin{document}


\title{Polarization-control of absorption of virtual dressed-states in helium}


\author{Maurizio~Reduzzi$^1$, Johan~Hummert$^{1}$, Antoine~Dubrouil$^1$, Francesca~Calegari$^2$, Mauro~Nisoli$^{1,2}$, Fabio~Frassetto$^{3}$, Luca~Poletto$^{3}$, Shaohao~Chen$^{4}$, Mengxi~Wu$^{4}$, Mette~B.~Gaarde$^{4}$, Kenneth~Schafer$^{4}$, Giuseppe~Sansone$^{1,2}$}

\affiliation{ 1) Dipartimento di Fisica, Politecnico Piazza Leonardo da Vinci 32, 20133 Milano Italy\\
2) Institute of Photonics and Nanotechnologies, CNR Politecnico, Piazza Leonardo da Vinci 32, 20133 Milano Italy\\
3) Institute of Photonics and Nanotechnologies, CNR via Trasea 7, 35131 Padova Italy\\
4) Department of Physics and Astronomy, Louisiana State University, Baton Rouge, LA 70803-4001, USA\\
}


\date{\today}

\begin{abstract}
The extreme ultraviolet absorption spectrum of an atom is strongly modified in the presence of a synchronized intense infrared field. In this work
we demonstrate control of the absorption properties of helium atoms dressed by an infrared pulse by changing the relative polarization of the infrared and extreme ultraviolet fields.
Light-induced features associated with the dressed $1s2s$, $1s3s$ and $1s3d$  states, referred to as $2s^{+}$, $3s^{\pm}$ and $3d^{\pm}$ light induced states, are shown to be strongly modified or even eliminated when the relative polarization is rotated. The experimental results agree well with calculations based on the solution of the time-dependent Schr\"{o}dinger equation using
a restricted excitation model that allows efficient treatment of the three dimensional problem.  We also present an analysis of the light induced states based on
Floquet theory, which allows for a simple explanation of their properties.
Our results open a new route to creating controllable superpositions of dipole allowed and non-dipole allowed states in atoms and molecules.
\end{abstract}


\maketitle

The interaction of light with matter is one of the most fundamental processes occurring in nature. Depending on the intensity of the electric field, different phenomena can occur ranging from the Stark shift of energy levels (for low electric field intensity) up to photoionization by tunneling (for intense electric fields). The few-femtosecond optical period of visible and near infrared radiation ($T_0=2.6$ fs for $\lambda=800$ nm) indicates that fundamental modifications of the electronic structure, due to the interaction with the external field, occur on a timescale of a few tens or hundreds of attoseconds \cite{RMP-Krausz-2009, PE-Nisoli-2009}. A recent example of modifications on this timescale is the measurement of the time-dependent dipole induced by a moderately strong infrared (IR) field~\cite{PRL-Neidel-2013}.
Furthermore, a signature of the sub-cycle Stark shifts induced by an intense IR field on the singly excited energy levels of helium was measured in the transmitted spectrum of an isolated extreme ultraviolet (XUV) attosecond pulse with varying delay between the two pulses~\cite{PRL-Chini-2012}.
This latter technique is usually named attosecond transient absorption spectroscopy and it has been used to investigate femtosecond and sub-cycle dynamics in noble gas atoms~\cite{PRL-Holler-2011, CP-Loh-2008, NJP-Lucchini-2013, OTT-arxiv, PRL-Wang-2010, PRA-Wang-2013, PRA-Bernhardt-2014}. With this technique it has been observed that the IR induces additional absorption features far from any transitions allowed by single photon absorption~\cite{PRA-Chen-2012, SR-Chini-2013}. These features can be linked to laser-dressed states and exhibit half cycle oscillations with changing delay between the IR and XUV fields.\\
In this work, we use attosecond transient absorption to investigate how the absorption of IR-dressed states can be controlled by varying the relative polarization of the attosecond and IR fields when they overlap in time. We demonstrate that  control of the relative polarization angle between the fields allows one to selectively excite particular groups of dressed states, depending on the quantum numbers of the undressed states that are coupled by the IR field.  We present two sets of calculations that agree well with the experimental observations. One is a numerical solution of the time-dependent Schr\"{o}dinger equation which agrees well with the experimental spectra over a wide frequency range. The other is a few level model based on Floquet theory,  A simple intuitive picture of the absorption can be gained by assuming that the attosecond XUV pulse creates a wave packet that is a superposition of IR-dressed excited states. The experimental results are then understood by considering the matrix elements of the relevant transitions between the undressed ground state and the manifold of dressed excited states. This approximation is justified because the moderate IR intensity in our experiment is  too low to excite or ionize the atom, but it appears as a strong field to electrons that are excited by the attosecond XUV pulse~\cite{PRA-Chen-2012, SR-Chini-2013}.

\section{Experiment}
\begin{figure}[htb]
\centering\includegraphics[width=10cm]{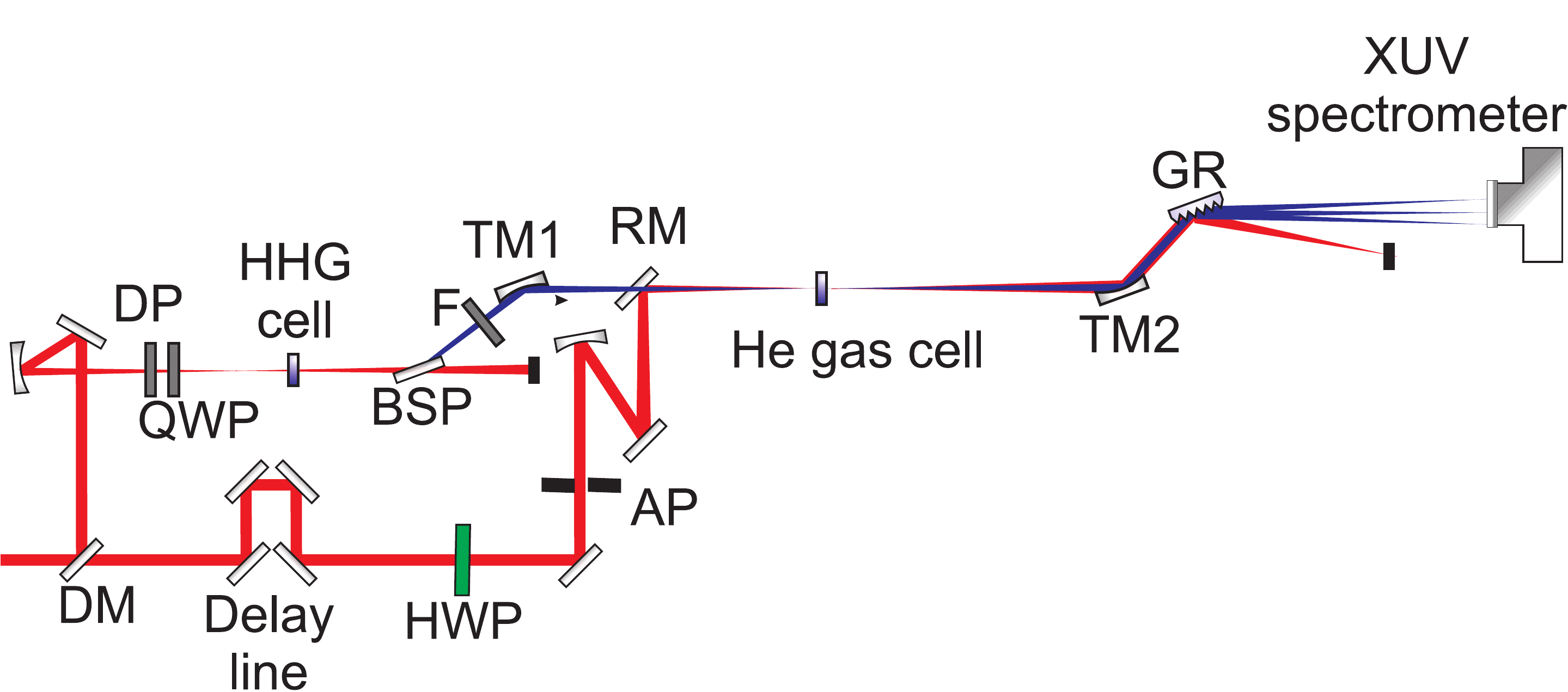}
\caption{Experimental setup. DM:~drilled mirror, DP:~delay plate, QWP:~quarter-wave plate, HWP:~half-wave plate, BSP:~beam-separator, F:~Filter, TM1/2:~toroidal mirrors, AP:~aperture, RM:~recombination mirror, GR:~grating.}
\label{Fig1}
\end{figure}

The experimental setup is shown schematically in Fig.~\ref{Fig1}. Few-cycle carrier-envelope phase (CEP) stabilized IR pulses were split using a drilled mirror (DM). The polarization of the central part (transmitted through the hole) was modulated in time using a delay plate (DP) and a quarter-wave plate (QWP) in order to allow the recollision of a single electronic wave packet with the parent ion, thus ensuring generation of an XUV continuum \cite{NATPHYS-SOLA-2006,SCIENCE-Sansone-2006}. The XUV radiation generated in the high-order harmonic generation (HHG) cell was reflected on a beam separator (BSP), which transmitted the p-component of the IR beam and efficiently reflected the XUV light, and by a toroidal mirror (TM1), which refocused the XUV radiation in a 3-mm-thick helium gas cell. After the beam separator, a 200-nm-thick aluminum filter (F) filtered the co-propagating IR radiation and the low-order harmonics. The spectral dispersion introduced by the birefringent plates (DP and QWP) was pre-compensated using a chirped mirror set at the output of the hollow fiber compressor (not shown).\\The annular beam reflected by the drilled mirror was delayed by a translation stage positioned on a piezo actuator. The annular beam passed through a half-wave plate (HWP), which rotated the polarization of the dressing pulse with respect to the polarization of the XUV field. The HWP allowed one to continuously change the relative angle $\theta$ between the polarizations of the attosecond radiation and the IR dressing pulse.
The dispersion was optimized to get the shortest pulses on the HHG arm for the generation of a single or a pair of attosecond pulses. However, the different dispersion introduced by the two birefringent plates on the HHG arm and by the half-wave plate on the dressing arm resulted in the non-optimal compression of the dressing pulse, which is retained a pedestal lasting a few tens of femtoseconds.
This low electric field component, however, does not influence the main conclusions that can be drawn from the experimental data. The intensity of the IR pulse was changed by using a variable aperture (AP). The XUV light and the dressing pulse were collinearly recombined using a second drilled mirror (RM). The XUV light transmitted through the helium gas cell was analyzed using a flat-field XUV spectrometer composed of a toroidal mirror (TM2) that refocuses the XUV radiation and a concave grating (GR) with 600~lines/mm that disperses the XUV components. The dispersed XUV light was finally collected onto a MCP coupled to a phosphor screen and a CCD-camera.\\
We performed measurements with and without CEP-stabilization of the driving pulses, without observing any remarkable difference in the delay-dependence of the helium absorption optical density. For CEP-stabilized pulses, a single attosecond pulse with a duration of 370~as was characterized using the FROG-CRAB method~\cite{PRA-Mairesse-2005}. In the case of CEP-unstabilized pulses, the XUV radiation was emitted as either one or two attosecond pulses \cite{PRA-Sansone-2009a, PRA-Sansone-2009b}. Since the emission of the attosecond bursts is linked to the IR field oscillations, the phase coherence between the attosecond pulses and the oscillation of the dressing pulse is maintained even without CEP stabilization. The measurements showed in the following were obtained without CEP stabilization.

%
%

\begin{figure}[htb]
\centering\includegraphics[width=6cm]{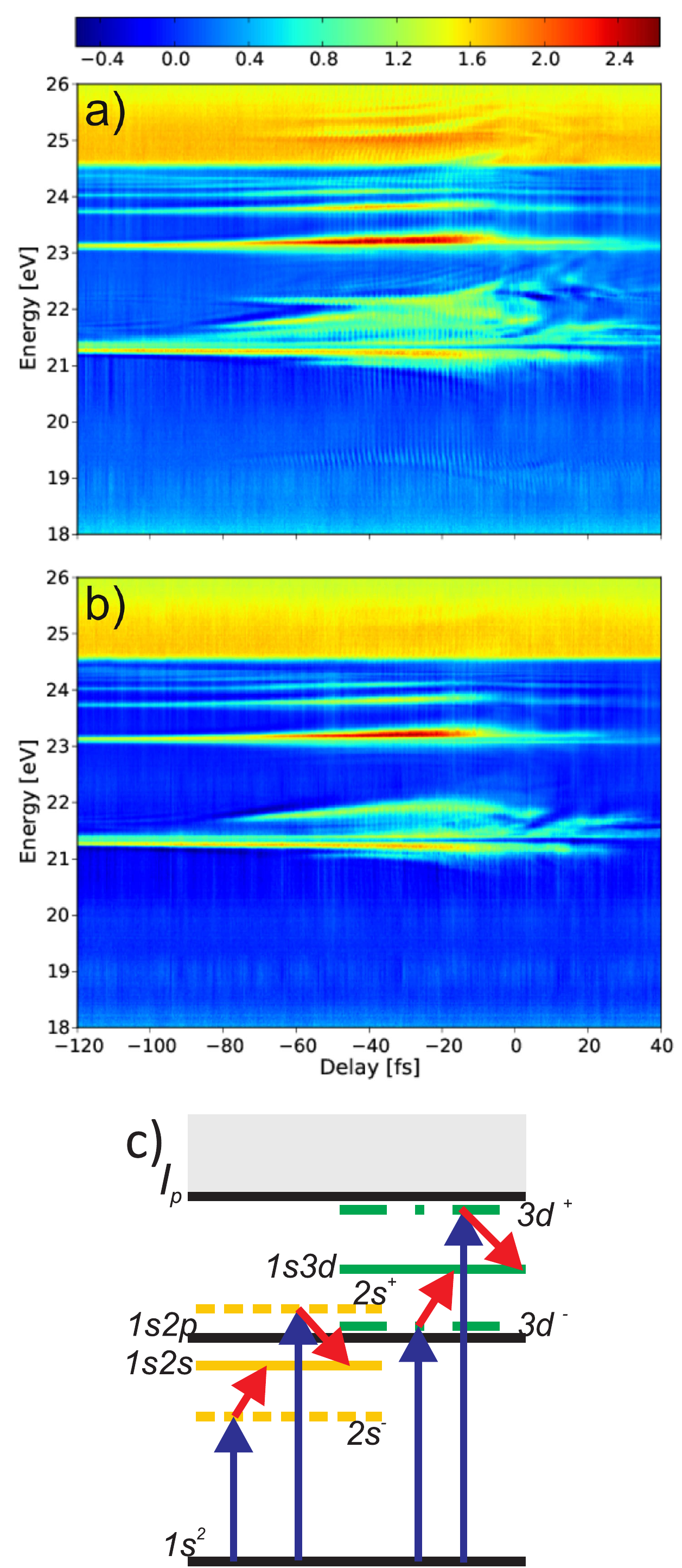}
\caption{Optical densities as a function of the relative delay between the XUV and IR pulses for parallel $\theta=0^{\circ}$ (a) and perpendicular $\theta=90^{\circ}$ (b) polarizations. Schematic representation of the two-color excitation pathways leading to the single-photon forbidden transitions to the unperturbed states $1s2s$ and $1s3d$ (solid lines), through the laser-dressed states $2s^{\pm}$ (dashed lines) and $3d^{\pm}$ (dot-dashed lines) (c). $I_p$ indicates the ionization potential.}
\label{Fig2}
\end{figure}

\begin{figure}[htb]
\centering\includegraphics[width=10cm]{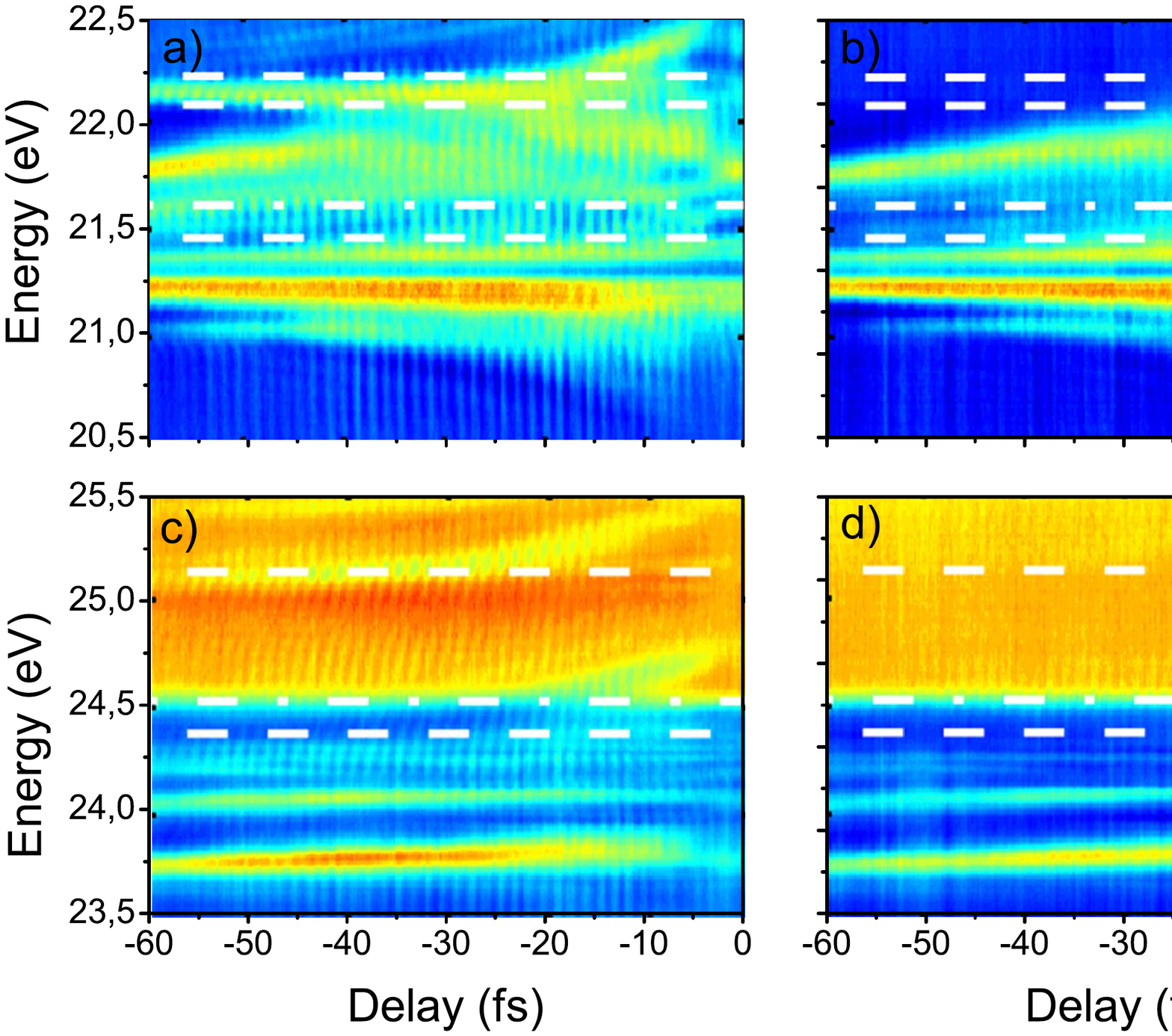}
\caption{Optical density as a function of the relative delay between the XUV and IR pulses in the energy ranges 20.5-22.5 eV (a,b) and 23.5-25.5 eV (c,d) for parallel $\theta=0^{\circ}$ (a,c) and perpendicular $\theta=90^{\circ}$ polarizations (b,d). The dashed lines indicate (from the lowest energy) the expected energies of the $ns^{\pm}$ dressed states $3s^-$, $2s^+$, and $4s^-$ (a,b) and $3s^+$ and $4s^+$ (c,d); the dash-dotted lines indicate the expected energy of the $3d^-$ (a,b) and $3d^+$ dressed state (c,d).}
\label{Fig3}
\end{figure}

Figures~\ref{Fig2}a,b show the measured optical density for two different relative orientations of the polarization axis of the two fields. The optical density $\mathrm{OD}(\omega,\tau)$ is shown as a function of the frequency $\omega$ and of the relative delay $\tau$ and is given by: $\mathrm{OD}(\omega,\tau)=-\log(S(\omega,\tau)/S_0(\omega))$
where $S(\omega,\tau)$ is the transmitted XUV spectra measured for the delay $\tau$ and $S_0(\omega)$ is the XUV spectrum measured without IR field. The IR dressing pulse arrives after the XUV pulse for negative delays. At large negative delays ($\tau=-120$~fs), the optical density is characterized by features corresponding to the singly excited $1snp$ states of helium. These lines are only partially visible for positive delays ($\tau=40$~fs) due to the small resonance width compared to the XUV spectrometer resolution ($\Delta E\simeq 30$~meV). For negative delays, the IR field ($\mathrm{I\simeq 10^{12} W/cm^2}$) can ionize  the populated excited states, leading to a reduction of the lifetime and, therefore, to an increase of the linewidth. Around the time overlap $-80~\mathrm{fs}<t<10~\mathrm{fs}$, the optical density presents a complex structure characterized by absorption lines that have been attributed to the absorption of XUV light by laser-dressed states, which result from the interaction between the atom and the intense IR field~\cite{PRA-Chen-2012, SR-Chini-2013}. In this picture, which we discuss in detail below, the dipole forbidden excited states such as the $1sns$ and $1snd$ levels can be populated  via a two-photon process involving the absorption of an XUV photon and the absorption (through the $ns^-$ and $nd^-$ dressed states) or emission (through the $ns^+$ and $nd^+$ dressed states) of an IR photon (as shown in Fig.~\ref{Fig2}c).
The energies of the one-IR-photon laser-dressed states relevant for this work are reported in table~\ref{Table1}. In the case of parallel polarizations shown in Fig.~\ref{Fig2}a and Fig.~\ref{Fig3}a,c, the absorption via the dressed states $ns^{\pm}$ is visible below ($3s^{-}$, $2s^{+}$, $4s^{-}$, $3s^{+}$) and above ($4s^{+}$, $5s^{+}$, $6s^{+}$) the ionization threshold of helium. The laser-dressed state $3d^{-}$ is also clearly visible. These observations are consistent with previously reported experimental results~\cite{CP-Loh-2008, PRA-Chen-2012, SR-Chini-2013, NJP-Lucchini-2013}.\\

The absorption of the laser-dressed states $ns^{\pm}$ is characterized by IR half-cycle oscillations, which are the results of the interference of the two pathways leading to the population of dipole forbidden states $ns$ as shown in Fig.~\ref{Fig2}c and discussed in detail below.
For the perpendicular case (Fig.~\ref{Fig2}b and Fig.~\ref{Fig3}b,d), the features corresponding to the $ns^{\pm}$ dressed states disappear, while the absorption line corresponding to the $3d^{-}$ state is still visible. In particular, the prominent $2s^{+}$ line has disappeared, and the continuum part of the spectrum is characterized by a uniform absorption without any additional time-dependent structure. The small offset between the measured($\simeq 21.75$ eV) and the expected ($\simeq 21.61$ eV) energy of the $3d^{-}$ state can be attributed to a cycle-averaged ponderomotive shift of the energy level. When the two polarizations are oriented at $45^{\circ}$, an intermediate situation between the parallel and perpendicular cases is realized (not shown). In this condition the absorption associated to the $ns^{\pm}$ dressed states is weaker than in the parallel case, resulting in a reduced visibility of the half-cycle oscillations.

In the remainder of this paper two theoretical analysis of the experimental data are presented. The first is a numerical solution of the time-dependent Schr\"{o}dinger equation (TDSE) in the single active electron (SAE) approximation. The second is an interpretation in terms of the dressed states of the excited atom in the IR laser field.

\begin{table}
\caption{\label{T  able1}Energy levels of singly excited, one-photon dressed states of helium for the IR-photon energy $\hbar\omega=1.46$ eV ($\lambda=850$ nm)}
\begin{tabular}{|l|l|l|}
\hline Energy Level	&Dressed States	&Energy [eV]	\\\hline
$1s2s$ [20.62 eV]	&$2s^-$	&19.16		\\
			&$2s^+$	&22.08		\\\hline
$1s3s$ [22.92 eV]	&$3s^-$	&21.46		\\
			&$3s^+$	&24.38		\\\hline
$1s3d$ [23.07 eV]	&$3d^-$	&21.61		\\
			&$3d^+$	&24.53		\\\hline
$1s4s$ [23.67 eV]	&$4s^-$	&22.21		\\
			&$4s^+$	&25.13		\\\hline
$1s5s$ [24.01 eV]	&$5s^-$	&22.55		\\
			&$5s^+$	&25.47		\\\hline
$1s6s$ [24.19 eV]	&$6s^-$	&22.73		\\
			&$6s^+$	&25.65		\\\hline
\end{tabular}
\label{Table1}
\end{table}

\section{Theory}
\subsection{Restricted Excitation Model}
 In this section the time-dependent Schr\"{o}dinger equation (TDSE) is solved numerically to study the effects of changing the laser polarization on transient absorption.
 To simplify the calculations we use a restricted excitation model (REM) \cite{Brabec-2008}.  In the REM the wave function is partitioned into ``ground state''  and  ``excited state'' pieces.
The weak attosecond XUV pulse is assumed only to excite amplitude out of the ground state and the IR pulse is assumed to act only on the
excited/continuum portion of the wave function. This is consistent  with the idea that while the IR field is too weak to excite the atom, once the system is excited the IR field acts in a non-perturbative manner.
When the XUV and IR fields have parallel polarizations, solving the  time-dependent Schr\"{o}dinger equation (TDSE) in the single active electron (SAE) approximation~\cite{PRA-Gaarde-2011} is reduced to a two dimensional problem due to cylindrical symmetry.
For non-collinear polarizations, however, solving the TDSE becomes a true three-dimensional problem.
The REM allows us to avoid the extra work involved in the increased dimensionality.

The REM starts from the TDSE
for an atom in two light fields,
\begin{eqnarray}
\label{tdse}
i \frac{\partial}{\partial t} \Psi(t) = \left[ H_A + H_L(t) + H_X(t) \right] \Psi(t),
\end{eqnarray}
where $H_A$ is the atomic Hamiltonian in the SAE approximation, $H_L(t)$ is the interaction with the IR laser field,
and $H_X(t)$ is the interaction with the XUV field. We use atomic units throughout.
The XUV field has a central frequency $\omega_X$
and its interaction can be written as
\begin{eqnarray}
\label{xuv}
H_X(t)=H_X^+(t) e^{-i \omega_X t} + H_X^-(t) e^{i \omega_X t}.
\end{eqnarray}
In the REM the total time-dependent wave function is
\begin{eqnarray}
\label{wp}
\Psi(t) = e^{-i E_0 t} \psi_0 + e^{-i \Delta t} \widetilde{\psi}(t),
\end{eqnarray}
where $E_0$ and $\psi_0$ are the ground state energy and wave function respectively,
and $\Delta=E_0+\omega_X$ is the detuning from threshold.
The full wave function is thus split into two pieces, a ground state portion that oscillates at
a frequency set by $E_0$, and an  ``excited'' portion that has been formed via one photon absorption and oscillates at approximately $\Delta$.

Substituting Eq.~(\ref{xuv}) and (\ref{wp}) into Eq.~(\ref{tdse}),
yields an equation for the time evolution of $\widetilde{\psi}(t)$ which is equivalent to
solving the full TDSE:
\begin{eqnarray}
\label{ees1}
i \frac{\partial}{\partial t} \widetilde{\psi}(t) &=& \left[e^{i \omega_X t} H_L(t) + H_X^+(t) + e^{2i\omega_X t} H_X^-(t) \right] \psi_0  \nonumber \\
 &+& \left[ H_A + H_L(t) + H_X(t) - \Delta \right] \widetilde{\psi}(t).
\end{eqnarray}
In light of the experimental conditions, {\it i.e.,} the modest IR intensity and the high XUV frequency, we can make the following  approximations:
(i)  the two rapidly varying terms that multiply $\psi_0$ are dropped while the more slowly varying term $H_X^+(t)$ is retained, and
(ii) the term $H_X(t)$ that acts on $\widetilde{\psi}(t)$ is dropped because the IR laser couples the excited states strongly while the XUV field does not.
With these approximations we obtain the REM equation for the wave packet $\widetilde{\psi}(t)$:
\begin{eqnarray}
\label{ees2}
i \frac{\partial}{\partial t} \widetilde{\psi}(t) = H_X^+(t) \psi_0 
 + \left[ H_A + H_L(t) - \Delta \right] \widetilde{\psi}(t)
\end{eqnarray}
This is an inhomogeneous equation where the term $H_X^+(t) \psi_0$ acts as a time-dependent source function.
This makes it convenient to deal with two light fields of different polarization directions.

We take both light fields to be linearly polarized and adopt the dipole approximation.
The laser-atom interactions are $H_L(t)= {\cal E}_L(t)\, \mathbf{u}_L \cdot \mathbf{r}$
and $H_X^+(t)=E_X^{+}(t)\, \mathbf{u}_X \cdot \mathbf{r}$,
where $\mathbf{u}_L$ and $\mathbf{u}_X$ are the polarization directions of the laser and XUV fields,
${\cal E}_L(t)$ is the full IR laser field and $E_X^{+}(t)$ is the complex time-dependent XUV field envelope.
We define a source wave function $\psi_s=\left(\mathbf{u}_X \cdot \mathbf{r}\right)\, \psi_0$
and further rewrite the first term on the right-hand side of Eq. (\ref{ees2}) as $E_X^{+}(t) \psi_s$. 
A short time step $\delta t$ is taken by calculating
\begin{equation}
\widetilde{\psi}(t+\delta t)= e^{-i(H_A+H_L-\Delta)\delta t}\left(\widetilde{\psi}(t)
-i\, \delta t \, E_X^{+}(t)\psi_s\right).
\end{equation}
At every time step where the XUV field is non-zero it adds a small amount to the excited wave packet. This source term has a distribution over the magnetic quantum number
$m$ that the laser field (polarized along the $z$ axis) then preserves during the time evolution from $t$ to $t+\delta t$.
This allows us to expand $\widetilde{\psi}(t)$ on a spherical harmonic basis with a fixed number of $m$ states
and solve Eq.~(\ref{ees2}) for each $m$ separately using the methods discussed in detail in~\cite{PRA-Gaarde-2011}.

In the following we consider the case of a helium atom interacting with either parallel or perpendicular polarizations of the pump and probe fields,
since within the REM other angles are just linear combinations of these two cases.
In the parallel case both fields are polarized in $\mathrm{z}$ direction,
thus the source wave function is
\begin{eqnarray}
\label{src1}
\psi_s = z \psi_0 =
\frac{1}{\sqrt{3}} r R_0 (r)Y_{1}^{0},
\end{eqnarray}
where $R_0$ is the helium ground state radial wave function in the  SAE and $Y_{1}^{0}$ is a spherical harmonic.
In this case the XUV field adds amplitude to the $l=1, m=0$ component of $\widetilde{\psi}(t)$ at every time step.
The IR field couples this to all other $l$ states with $m=0$ during the time evolution.
In the orthogonal case, the XUV field is polarized in $x$ direction
and the source wave function is calculated as
\begin{eqnarray}
\label{src2}
\psi_s = x \psi_0
= -\frac{1}{\sqrt{6}}\, r R_0 \left( Y_{1}^{1} - Y_{1}^{-1} \right).
\end{eqnarray}
In this case the XUV field adds amplitude to the \mbox{$l=1, m=\pm1$} components of $\widetilde{\psi}(t)$,
and the IR field couples this to other allowed $l$ states but preserves $m=\pm1$ during the time step.
Note that the $l=0$ states are never populated in the orthogonal case. Following these considerations,
the excited wave packet is expanded in a mixed radial grid-spherical harmonic basis as
\begin{equation}
\widetilde{\psi}(r,\theta,\phi,t)=\sum_{m=-1}^{m=+1}\sum_{l}\widetilde{R}_{l}^{m}(r,t) Y_{l}^{m}(\theta,\phi)
\end{equation}

After solving the TDSE the transient absorption spectrum can be described at the single atom level by a frequency-dependent response function \cite{PRA-Gaarde-2011}:
\begin{eqnarray}
\label{resp}
\tilde S(\omega)= 2\ {\rm Im} \left[\tilde d(\omega) \tilde {\cal E}_X^* (\omega) \right],
\end{eqnarray}
where $\tilde d(\omega)$ and $\tilde {\cal E}_X (\omega)$ are the Fourier transforms of
the dipole moment $d(t)$ and the XUV field ${\cal E}_X (t)$ respectively.
Because we are only concerned with the dipole oscillation in the frequency range around $\omega_X$,
the dipole moment can be calculated for parallel polarizations as
\begin{eqnarray}
\label{dip1}
d(t) & = & e^{-i \omega_X t} < \psi_0 | z | \tilde{\psi}(t) >  + \, \mbox{c.c.}\\ \nonumber
     & = & \frac{1}{\sqrt{3}}\, e^{-i \omega_X t} \int dr R_0\, r\, \widetilde{R}_1^0(t) + \, \mbox{c.c.},
\end{eqnarray}
and for perpendicular polarizations as
\begin{eqnarray}
\label{dip2}
d(t) & = & e^{-i \omega_X t} < \psi_0 | x | \tilde{\psi}(t) >  + \, \mbox{c.c} \\ \nonumber
     & = & -\frac{1}{\sqrt{6}} e^{-i \omega_X t} \int dr R_0\, r \left( \widetilde{R}_1^{1}(t) - \widetilde{R}_1^{-1}(t) \right) + \, \mbox{c.c.}
\end{eqnarray}
Because of the fixed phase relationship between the $m=+1$ and $m=-1$ components of the wave function
we need only calculate the $m=1$ component and multiply its dipole contribution by two.


For the detailed simulations we use a laser field of the form ${\cal E}_L(t)={\cal E}_0 \cos^2(\pi t/T_L)\sin(\omega_L t)$ where ${\cal E}_0$ is the IR  field strength
and the pulse lasts from $t=-T_L/2$ to $t=+T_L/2$, and $E_X^{+}(t)=E_0\cos^2[\pi (t-\tau)/T_X]$ is the time-dependent envelope of the XUV pulse.
The time delay between the two pulses is $\tau$. The central peak of the 5 cycle full width at half maximum (FWHM) IR pulse is 800 nm and the peak intensity is $3 \times 10^{12}$ W/cm$^2$. The XUV pulse has a central frequency of 25 eV, a peak intensity of $1 \times 10^{11}$ W/cm$^2$ and a FWHM duration of 300 as.

\begin{figure}[htb]
\centering\includegraphics[width=7.5cm]{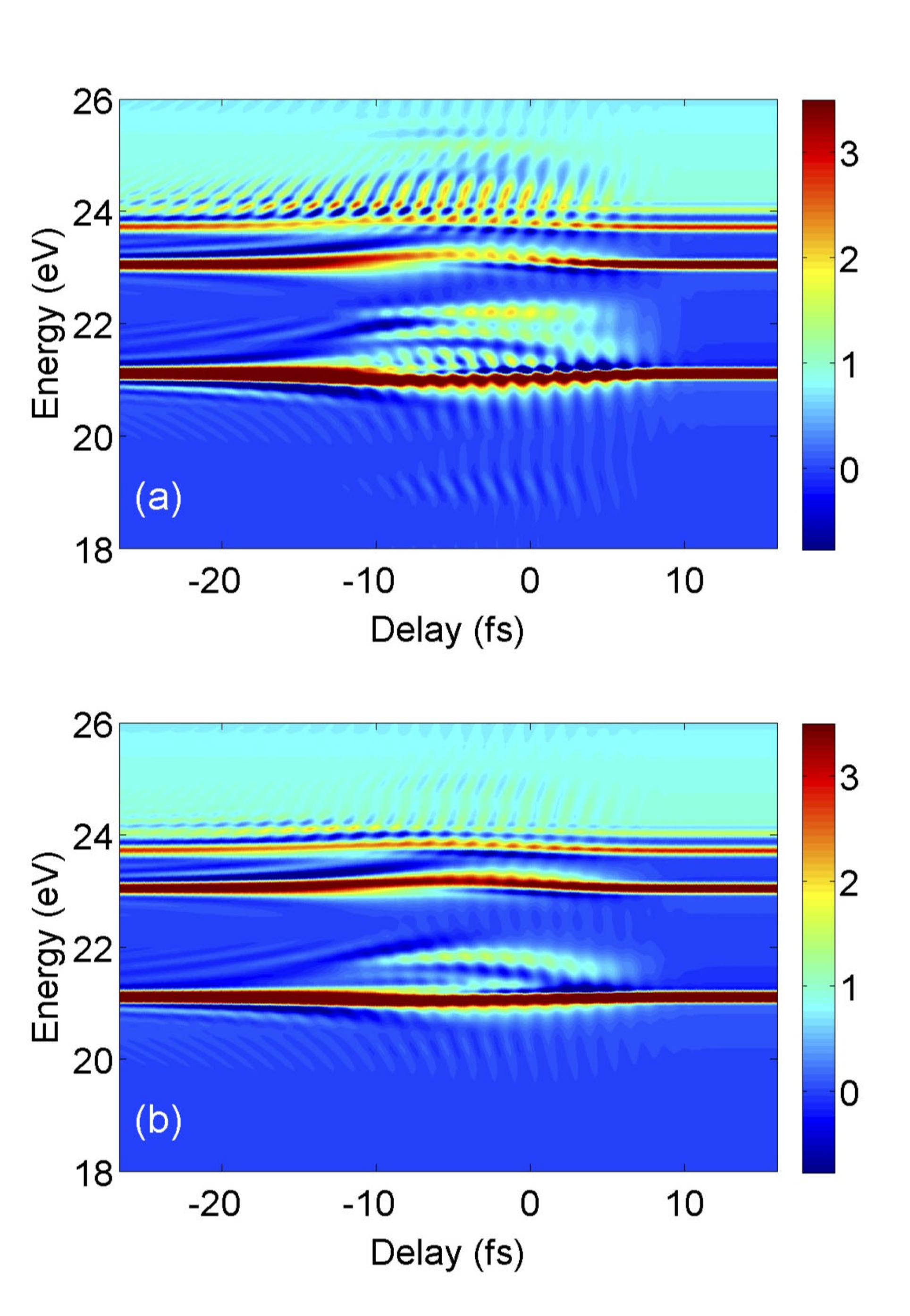}
\caption{Calculated single atom helium response (Eq.~\ref{resp}) as a function of the relative delay between the XUV and IR pulses for parallel $\theta=0^{\circ}$ (a) and perpendicular $\theta=90^{\circ}$(b) polarizations.}
\label{Fig4}
\end{figure}

The results of the simulations are shown in Fig.~\ref{Fig4}a,b for the parallel and perpendicular case, respectively. We concentrate on the three absorption features between the field free $2p$ and $3p$ states (21.1 to 23.0 eV) that are visible when the two pulses overlap in time (delays between -10 and +10 fs) and the XUV and IR polarizations are parallel.  These are the $3s^-$, $2s^+$ and $3d^-$ dressed states \cite{PRA-Chen-2012} discussed in connection with the experimental data shown in Fig.~3. As in the experiment, the highest and lowest of these three features, corresponding to $2s^+$ and $3s^{-}$ dressed states disappear when the polarizations are perpendicular. The absorption line at 21.9 eV, corresponding to the $3d^{-}$ dressed state  is, however, visible for both polarizations. The absorption features above the ionization energy are greatly diminished in the calculations, in agreement with the experimental findings. The simulations, therefore, reproduce all main features observed in the experimental data. We turn now to a discussion of these results based on a simple Floquet picture of XUV absorption in the overlap region of the delay scan.

\subsection{Floquet picture}
In order to gain a simple physical picture of the differences observed for parallel and perpendicular polarizations, it is useful to consider the Floquet representation of the helium system dressed by the IR field and interacting with the XUV pulse. In this description, the state describing the atom is considered as a coherent superposition of laser-dressed states, which can be ordered as a function of the number of IR photons involved. The main approximation that we employ is again the idea that the ground state is not affected by the IR laser and so the transient absorption measurement probes  transitions between the undressed ground state and the manifold of excited states dressed by the IR field. We can then consider the matrix element between the ground state and the excited-dressed states in order to evidence differences between the absorption in the parallel and perpendicular case.

Before discussing the more complicated case of the full helium atom, we first consider a simple model for the feature we have labeled as the $2s^{+}$ light induced state. We simplify the helium atom to a three state system consisting of the ground $1s$ state and the $2s$ and $2p$ excited states. The energies and dipole couplings are the same as in the SAE calculations. We use the same laser parameters as in the full simulation above, except that we move the central wavelength of the XUV to the $1s-2p$ energy to better highlight the light-induced features. The result of a full solution of the three-level TDSE is shown in Fig.~5. The absorption features appear near the $2p$ energy and also at
energies that are approximately one IR photon away from the $2s$ state and two photons away from the $2p$ state. There are also visible half cycle oscillations as in the experiment and the full TDSE simulations. We now discuss how the three level model can be explained in terms of Floquet states. We use the same notation as \cite{PRA-Tong-2010} where
laser-dressed photo absorption in helium was analyzed in terms of Floquet theory.
%
%
\begin{figure}[t]
\centering\includegraphics[width=8cm]{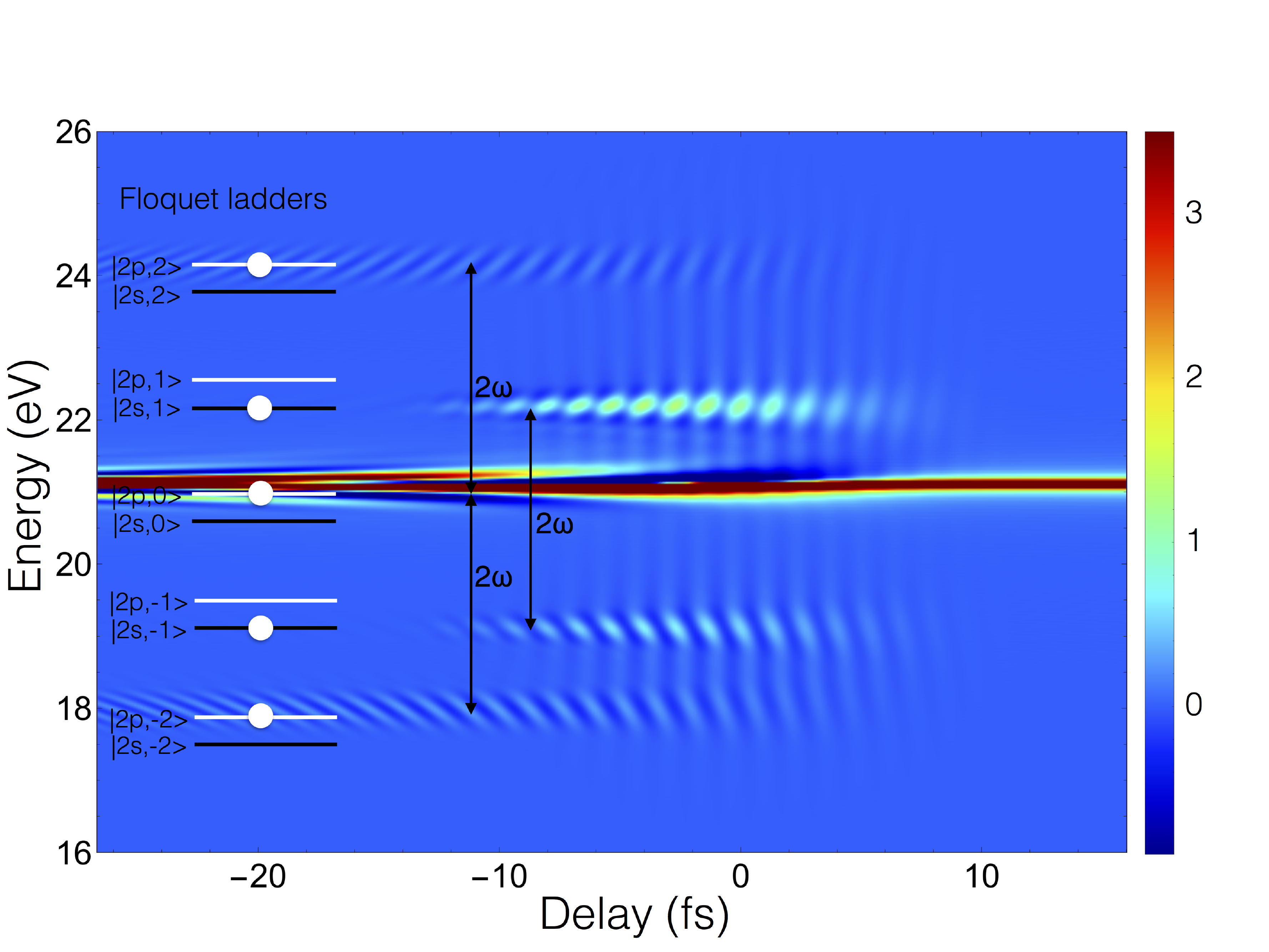}
\caption{Calculated single atom response (Eq.~\ref{resp})for the three level helium atom. On the left are shown the
positions of the dressed states $|\alpha,n\rangle$ where $\alpha$ is either $2s$ or $2p$ depending on the state in zero field. The white dots
show states that have a non-zero transition moment to the $1s$ ground state. }
\label{Fig5}
\end{figure}

In the three level model, the Floquet states that describe the dressed excited  states are built from products of the $2p$ or $2s$ states with an integer number of photons. Diagonalizing the two state plus laser field Floquet Hamiltonian~\cite{PR-Chu-2004} for a constant IR intensity yields time-independent states $|\phi_{\alpha,n}\rangle$, where the $\alpha$ label refers to the state in zero field and is either $2s$ or $2p$ in our model, and $n$ is the index for photon numbers involved. Each atomic state $\alpha$ gives rise to a ``ladder" of states with energies $\epsilon_{\alpha}+n\omega$.
These energies are shown on the left hand side of Fig.~5 and are calculated using the same peak intensity as the calculations shown in Fig. 4. The Floquet energies
$\epsilon_{\alpha}$ are close to the field free $2s$ and $2p$ energies in this case because of the large detuning ($\sim$~1~eV) of the $2p-2s$ energy from $\omega$. For smaller detunings the energies can differ substantially.

 As discussed in ~\cite{PRA-Tong-2010} the Floquet ladder states can be used to construct time-dependent Floquet states $|\Psi_\alpha(t,\tau)\rangle$ that can then be used to describe the time-evolution of the system.  Because they are not coupled by the IR field, the time evolution of a Floquet state following excitation at $t=0$ is given by
\begin{equation}
| \Psi_\alpha(t,\tau)\rangle=e^{-i\epsilon_\alpha t}\sum_n e^{-in\omega (t+\tau)}|\phi_{\alpha,n}\rangle,
\label{eqn:Floquet_states}
\end{equation}
where we have accounted for the delay $\tau$ between the XUV pulse and the IR field by adding a phase $e^{-i n \omega \tau}$ to each of the ladder states.

Using the  time-dependent Floquet states $|\Psi_{\alpha}\rangle$ as a basis, the time-dependent dipole moment can be calculated as follows. If we ignore the XUV pulse duration,  the initial  excited state wave function is proportional to $\hat{\mu}_X|\psi_0\rangle$  where $\hat{\mu}_X$ is the dipole operator of the XUV field and $|\psi_0\rangle$ is the ground state. This can be expressed as a  superposition of the Floquet states at $t=0$:
\begin{equation}
|\Psi(t=0,\tau)\rangle=
\sum_\alpha C_\alpha^\tau\, |\Psi_\alpha(t=0,\tau)\rangle,
\end{equation}
where $C_\alpha^\tau$ is the dipole transition element from the ground state to the different Floquet states,
\begin{eqnarray}
\label{eqn:Floquet_dipole_matrix}
C_\alpha^\tau &=& \langle\Psi_\alpha(t=0,\tau) | \hat\mu_X | \psi_0\rangle \nonumber \\[1em]
&=&\sum_m e^{i m \omega\tau} \langle\phi_{\alpha,m}|\hat\mu_X|\psi_0\rangle.
\end{eqnarray}
Once the excited state wave function has been decomposed at $t=0$ into dressed states, it can be found at any time $t>0$ using Eq.~\ref{eqn:Floquet_states}:
\begin{eqnarray}
| \Psi(t,\tau)\rangle &=& \sum_\alpha C_\alpha^\tau \, |\Psi_\alpha(t,\tau)\rangle.
\end{eqnarray}
This expression is equivalent to what is called $\widetilde{\psi}(t)$ in Eq.~\ref{wp}. Using it to construct the time dependent dipole as  in Eq.~\ref{dip1}
gives
\begin{widetext}
\begin{equation}
\label{eqn:Floquet_dipole}
d(t,\tau) =\\
\sum_{\alpha,m,n}e^{-i (\epsilon_\alpha+m \omega-E_0) t} e^{i (n-m) \omega \tau} \langle \phi_{\alpha,n}|\hat\mu_X|\psi_0 \rangle \langle \psi_0|\hat\mu_X|\phi_{\alpha,m}\rangle + \mbox{c.c.},
\end{equation}
\end{widetext}
where $E_0$ is again the ground state energy. This equation, valid for short XUV pulse excitation of a system strongly coupled by an IR field,  is the main result of this section.
It  can explain many of the general features in the delay dependent absorption spectrum when the pulses overlap.

 To begin with, Eq.~\ref{eqn:Floquet_dipole} shows that the dipole moment oscillates at frequencies $\epsilon_\alpha+m \omega-E_0$, which means there can be absorption at those frequencies if they are present in the XUV spectrum, and if the corresponding dressed state has a non-zero transition moment to the ground state.  Due to  parity conservation, the XUV pulse can only populate every other state in each Floquet ladder, and in Fig.~5 we have put a white dot on the states in each ladder
that have a non-zero transition moment to the ground state. The absorption is also modulated as a function of delay at frequency $(n-m)\omega\tau$. Since parity considerations dictate that  $m-n$ must be an even number we expect to see oscillations in the absorption as a function of delay with a period of $T_0/2$, $T_0/4$, etc., with the half cycle ($2\omega$) oscillations being the strongest.

The light-induced states we have referred to as $2s^{+}$ and $2s^{-}$ are seen to be simply the $|2s,\pm 1\rangle$ dressed states. The amplitude of the light-induced features in $d(t,\tau)$ depend on the IR field through the matrix elements
$\langle \phi_{\alpha,n}|\hat{\mu}_X|\psi_0\rangle$, and they obviously last only until the IR field ends and the ladders collapse to the field free states. This explains why these features are broadened at positive delays. We note that absorption at the $|2p, -2\rangle$ energy can be seen for negative delays where the XUV and IR pulses do not overlap. In this case the
dressed state is populated by the turn on of the IR pulse when $2p$ population is redistributed over the $2p$ dressed states. In our experiment the XUV bandwidth does not
extend to this frequency, but this feature has been observed previously~\cite{SR-Chini-2013}.

From Eq.~\ref{eqn:Floquet_dipole} it can be seen that the $2\omega$ oscillations in the light-induced features are a direct indication of the sub-cycle duration of the XUV pulse. That is, the broad bandwidth of the XUV pulse coherently populates multiple states in each Floquet ladder so that the phases accumulated on those states can interfere and give delay dependent oscillations in the absorption spectrum. This conclusion is unchanged if one considers a short train of XUV pulses that are individually sub-cycle in duration, provided they are separated by $T_0/2$.
If we do not ignore the XUV pulse dutation, as we have been, then  Eq.~\ref{eqn:Floquet_dipole} can be generalized by  convoluting it with the XUV envelope
\begin{equation}
d(t,\tau)=\int d(t,\tau')\mathcal E_X(\tau')d\tau',
\end{equation}
and it can be easily shown that the delay dependent oscillations are strongly suppressed if the XUV pulse duration is longer than $T_0/2$.

The three level model can be extended to include more atomic states in the basis, leading to more Floquet ladders. The most important couplings
for the $2s$, $3s$ and $3d$ states are to the $2p$ and $3p$ states. The $n= \pm 1$ sideband of these dark states will be visible in the absorption spectrum near an energy
$\epsilon_{\alpha} \pm \omega$ if the XUV bandwidth overlaps their position. Because the IR pulse is always along the $z$ axis the dipole interaction $H_L(t)= {\cal E}_L(t) z$ couples states with $\Delta m=0$ only. This means that the Floquet states created by the IR field can be sorted into manifolds labeled by $m$. Since the ground state retains its $\ell=0, m=0$ character simple dipole selection rules dictate that the XUV interaction $H_X^{+}$ will only couple the ground state to dressed states that have some $\ell=1,m=0$ component when the XUV field is polarized parallel to the IR field, and will only couple the ground state to dressed states with some amount of  $\ell=1,m=\pm 1$ character when it is polarized orthogonal to the IR field. The transition element  (Eq.~\ref{eqn:Floquet_dipole_matrix}) between the ground $1s$ state and a dressed state,
\begin{equation}
C_{\alpha}^\tau=E_X^+<\Psi_{\alpha}(\mathbf{r},t=0,\tau)|\mathbf{u}_X\cdot\mathbf{r}|\psi_{1s}(\mathbf{r})>
\label{groundstate}
\end{equation}
reduces to the sum over contributions whose angular part can be expressed as:
\begin{eqnarray}
<Y_l^m|z|Y_{0}^0>&\propto&\delta_{l,1}\delta_{m,0}\quad\qquad\quad~\,\mathrm{parall.~case} \nonumber\\
<Y_l^m|x|Y_{0}^0>&\propto&\delta_{l,1}[\delta_{m,-1}-\delta_{m,1}]~\mathrm{perp.~case}
\label{Eqtransition2}
\end{eqnarray}
Thus we expect that, beginning with parallel XUV and IR polarizations, the $m=0$ dressed states will disappear from the absorption spectrum as the polarization is rotated, while $m=\pm 1$ states will begin to appear. In the extreme cases, only $m=0$ states appear for parallel polarization, and only $m=\pm 1$ states appear for perpendicular polarization.
This is consistent with the experiment results: the $2s^+$ light induced state with only $m=0$ character is strongest for parallel polarizations and vanishes for  perpendicular polarization, while the $3d^-$ state with both $m=0$ and $m=\pm 1$ character persists in both parallel and perpendicular cases.

\section{Conclusions}
In conclusion, we have investigated the polarization dependence of the absorption lines of excited states of helium dressed by an IR pulse. The experimental data evidence a clear difference in the absorption features of $ns^{\pm}$ and $nd^{\pm}$ dressed state between the parallel and perpendicular polarization cases. The different behavior is confirmed by TDSE simulations and can be interpreted in terms of the angular part of the integral contributing to the dressing of the excited states by the IR field.
These observations indicate that the IR field can be used to create a coherent superposition of states (ground state $1s$ and $3d$ state for example) characterized by a non-uniform population of $m-$quantum number states. In the time domain, this superposition corresponds to a charge breathing of few hundreds of attosecond (due to the energy spacing between the ground and excited states) characterized by a symmetry that can be controlled by the polarization of the IR field.\\

\begin{acknowledgments}
Financial support by the Alexander von Humboldt Foundation (Project "Tirinto"), the Italian Ministry of Research (Project FIRB  No. RBID08CRXK), the Marie Curie Research Training Network ATTOFEL, and the European Research Council under the European Community's Seventh Framework Programme (FP7/2007-2013) / ERC grant agreement n. 227355 - ELYCHE is gratefully acknowledged. This work was supported by the Director, Office of Science, Office of Basic Energy Sciences, and by the Division of Chemical Sciences, Geosciences, and Biosciences of the U.S. Department of Energy under Contract No. DE-FG02-13ER16403. High-performance computing resources were provided by the Louisiana Optical Network Initiative (LONI).
\end{acknowledgments}
e-mail: giuseppe.sansone@polimi.it\\ e-mail: schafer@phys.lsu.edu
\bibliography{PRA_Helium_bib}
\end{document}